\documentclass[useAMS,usenatbib]{mn2e}

\newcommand{\tild}{\raise.27ex\hbox{$\scriptstyle\sim$}}
\usepackage{graphicx}
\usepackage{overpic}
\usepackage{color}

\newcommand{\Cassini}{\textit{Cassini~}}
\newcommand{\CassiniVIMS}{\textit{Cassini}-VIMS~}

\usepackage{aas_macros}

\newcommand{\executeiffilenewer}[3]{%
 \ifnum\pdfstrcmp{\pdffilemoddate{#1}}%
 {\pdffilemoddate{#2}}>0%
 {\immediate\write18{#3}}\fi%
}

\title[The weather report from IRC+10216]{The weather report from IRC+10216: Evolving irregular clouds envelop carbon star}
\author[Stewart et al.]
{\parbox{\textwidth}{P. N. Stewart$^{1}$\thanks{E-mail:p.stewart@physics.usyd.edu.au (PNS)},
P. G. Tuthill$^{1}$,
J. D. Monnier$^{2}$,
M. J. Ireland$^{3}$,
M. M. Hedman$^{4}$,
P. D. Nicholson$^{5}$,
and S. Lacour$^{6}$}\vspace{0.4cm}\\
$^{1}$Sydney Institute for Astronomy, School of Physics, The University of Sydney, NSW 2006, Australia\\
$^{2}$Astronomy Department, University of Michigan (Astronomy), 500 Church St, Ann Arbor, MI 48109, USA\\
$^{3}$Research School of Astronomy \& Astrophysics, Australian National University, Canberra, ACT 2611, Australia \\
$^{4}$Department of Physics, University of Idaho, Moscow, ID 83844, USA\\
$^{5}$Department of Astronomy, Cornell University, Ithaca, NY 14853, USA\\
$^{6}$LESIA, CNRS/UMR-8109, Observatoire de Paris, UPMC, Université Paris Diderot, 5 place Jules Janssen, Meudon, France}
\begin{document}

\date{DRAFT \today}

\pagerange{\pageref{firstpage}--\pageref{lastpage}} \pubyear{2015}

\maketitle

\label{firstpage}

\begin{abstract}
High angular resolution images of IRC+10216 are presented in several near infrared wavelengths spanning more than 8 years.
These maps have been reconstructed from interferometric observations obtained at both Keck and the VLT, and also from stellar occultations by the rings of Saturn observed with the \Cassini spacecraft.
The dynamic inner regions of the circumstellar environment are monitored over eight epochs ranging between January 2000 and July 2008.
The system is shown to experience substantial evolution within this period including the fading of many previously reported persistent features, some of which had been identified as the stellar photosphere.
These changes are discussed in context of existing models for the nature of the underlying star and the circumstellar environment.
With access to these new images, we are able to report that none of the previously identified bright spots in fact contain the star, which is buried in its own dust and not directly visible in the near infrared.
\end{abstract}


\section{Introduction}

One of the most extensively studied evolved stars is the carbon-rich, IRC+10216.
Also known as CW Leo, it is a long period variable on the Asymptotic Giant Branch (AGB).
It is believed to be on the cusp of planetary nebula formation, a process which is potentially already under way.
IRC+10216 is known to be embedded in, and strongly extincted by, an expanding shroud of material originating from the star itself.

The circumstellar environment of IRC+10216 has long been known to be complex and continually evolving.
As the infrared-brightest example of an AGB star experiencing heavy mass-loss, it has been exhaustively studied in many wavelengths with diverse observational techniques.
Shell-like structures have been previously detected beyond 1 arc-second from the star.
\citet{Mauron1999, Mauron2000} discovered evidence for shells out to \tild 50 arc-seconds within a 200 arc-second envelope, and~\citet{Decin2011} detected non-concentric shell-like arcs out to 320 arc-seconds.
These shells are shown to be irregularly separated and have a non-uniform density distribution, suggesting they have an irregular and asymmetric origin.
Far beyond this a very large 1280\,mas HI shell has been identified by~\citet{Matthews2015}, revealing the interactions of the stellar wind with the ISM.
Combining observations of the outer regions with those of the inner regions, ~\citet{Leao2006} studied the complex interaction between winds at different distances to help understand the mass-loss history of the star.

\begin{figure}
 \centering
 \def\svgwidth{1.\columnwidth}
 \executeiffilenewer{schematic.svg}{schematic.pdf}%
 {inkscape -z -D --file=schematic.svg %
 --export-pdf=schematic.pdf --export-latex}%
 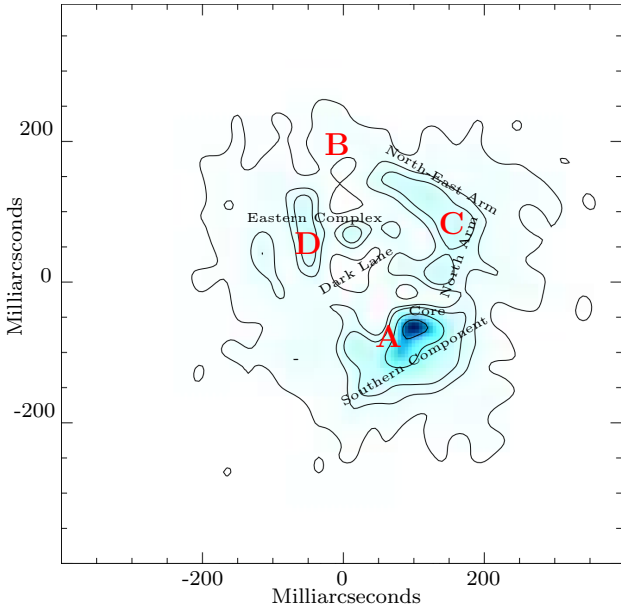%

 \caption{A Schematic showing the approximate positions of previously identified features in the nebula. The large red `A', `B', `C', and `D' are the~\citet{Weigelt1998}/\citet{Haniff1998} naming scheme.
 The smaller labels were the names used by~\citet{Tuthill2000} to identify structures. 
 The background map on which these are overlayed is from the first epoch of this campaign using NIRC's CH$_4$ filter.}
 \label{fig:schematic}
\end{figure}

The inner regions of the system have been imaged in the infrared many times over the past two decades in attempts to identify the location of the star itself, and to refine radiative transfer models describing its mass-loss.
The first detection of asymmetry in the inner circumstellar region of the star was made by~\citet{Kastner1994} who found that this region appeared to have a bipolar structure.
Subsequently using a variety of techniques fine structure was detected on sub-arc-second scales which broke this axial symmetry.
Bright features within this structure were classified by~\citet{Weigelt1998} and~\citet{Haniff1998} in order of decreasing brightness as \textit{A}, \textit{B}, \textit{C} and \textit{D} as indicated in Figure~\ref{fig:schematic}.
These labels were subsequently used by other authors as the evolution of the system was monitored.
The initial assertion that the brightest feature (\textit{A}) included the star itself was supported by~\citet{Tuthill2000, Tuthill2005} and~\citet{Richichi2003}.
This conclusion was contested in favour of the star being within feature \textit{B} by~\citet{Osterbart2000} and~\citet{Weigelt2002}, sparking a debate that continued for many years.
The star driving the winds and forming these structures is expected to exhibit a uniform disc angular diameter of around 29\,mas~\citep{Menten2012}.

Modelling the mass-loss behaviour of carbon stars has traditionally involved one dimensional radial models of the star.
These are inherently unable of producing the kinds of asymmetric structure which has been observed.
There have been some attempts to model such stars in 2D and 3D which have shown some promise~\citep{Woitke2006a,Freytag2008}.

Here we present high resolution observations of the inner regions of IRC+10216 from eight epochs spanning more than eight years, in several near-IR bands.
These observations reveal the continued evolution of this interesting object.
These data exhibit dramatic changes, contradicting most previous models which had interpreted the brightness distribution in terms of underlying structural elements.
Specifically we exclude previous identifications of the stellar photosphere, and find no persistent evidence in support of bipolarity.

\section{Observations}

The imagery presented herein was obtained using two complimentary techniques: aperture masking interferometry and kronocyclic tomography.
The latter is described in~\citet{Stewart2015a} and entails image recovery from \Cassini observations of stellar occultations by the Saturnian rings.
A summary of all observations used in this paper including filters, aperture masks, dates and instruments can be found in Table~\ref{tab:obs}.
The spectral specifications of the observations, including the filters used, are listed in Table~\ref{tab:filters}.

\begin{table}
  \caption{\label{tab:obs}Basic observational parameters.
  Observations using kronocyclic tomography have an empty mask column.
  }
  \centering
  \begin{tabular}{l c c l l}
\hline \hline
&Date&Instrument&Filter&Mask\\
\hline
1&25 Jan 2000&NIRC&CH$_4$&Annulus\\
&&&CH$_4$&21 Hole\\
&&&PAHCS&Annulus\\
&&&PAHCS&21 Hole\\

2&24 Jun 2000&NIRC&CH$_4$&21 Hole\\
&&&PAHCS&21 Hole\\

3&11 Jun 2001&NIRC&CH$_4$&Annulus\\
&&&CH$_4$&21 Hole\\
&&&PAHCS&Annulus\\

4&12 May 2003&NIRC&CH$_4$&Annulus\\
&&&CH$_4$&21 Hole\\
&&&PAHCS&Annulus\\

5&28 May 2004&NIRC&CH$_4$&Annulus\\
&&&CH$_4$&21 Hole\\
&&&PAHCS&Annulus\\

6&25 May 2005&NIRC&CH$_4$&Annulus\\
&&&CH$_4$&21 Hole\\
&&&PAHCS&Annulus\\
&&&PAHCS&21 Hole\\

7&15 Mar 2008&CONICA&IB\_2.12&18 Hole\\
&&&NB\_3.74&18 Hole\\
&&&NB\_4.05&18 Hole\\

8& Jun/Jul 2008&VIMS&2.66&\textit{}\\
&&&3.32&\textit{}\\
&&&3.99&\textit{}\\
&&&4.66&\textit{}\\

\end{tabular}

\end{table}

\begin{table}
  \caption{\label{tab:filters}Spectral Specifications.
  NIRC and CONICA observations utilised standard instrument filters whilst adjacent VIMS spectral channels were co-added to produce images from broad spectral bands.}
  \centering
  \begin{tabular}{c c c c}
\hline \hline
Instrument&Filter&$\lambda_c$&$\Delta \lambda$\\
&Name&($\mu m$)&($\mu m$)\\
\hline
NIRC&CH$_4$&2.269&0.155\\
&PAHCS&3.083&0.101\\
CONCIA&IB\_2.12&2.12&0.06\\
&NB\_3.14&3.740&0.02\\
&NB\_4.05&4.051&0.02\\

\CassiniVIMS&VIMS 2.66&2.66&0.66\\
&VIMS 3.32&3.32&0.67\\
&VIMS 3.99&3.99&0.67\\
&VIMS 4.66&4.66&0.69\\
\end{tabular}

\end{table}

The aperture masking imagery, based on~\citet{Tuthill2000} and~\citet{Tuthill2005}, is used to image IRC+10216 in several near-IR wavelengths.
The observations were made both with the 10\,m Keck I telescope and the 8\,m UT4 of the VLT using the NIRC (Near Infrared Camera) and CONICA (Coude Near Infrared Camera) instruments respectively.
Both the 21 hole NIRC aperture mask and the 18 hole CONICA aperture mask are 2D non-redundant patterns.
The geometries of both masks used with NIRC are described in~\citet{Tuthill2000a} and the mask used with CONICA is described in~\citet{Tuthill2010a} and~\citet{Lacour2011}.
The maps recovered from aperture masking observations are the noise weighted averages for all reconstructed images with a single filter in a single epoch.
The image reconstructions presented here have been performed with BSMem~\citep{Buscher1994} for NIRC observations and Mira~\citep{Thiebaut2008} for the CONICA epoch.
Maps produced by these algorithms were cross-checked against those produced with other means (VLBMEM~\citep{Sivia1987} and MACIM~\citep{Ireland2006b}) and found to produce consistent structure whilst maintaining a relatively low background noise level.
The aperture masking observations were a continuation of the programs undertaken by~\citet{Tuthill2000, Tuthill2005} and comprise seven previously unpublished epochs, between January 2000 and March 2008.

The images recovered using kronocyclic tomography are from \Cassini observations of stellar occultations in which the rings of Saturn pass in front of the target star.
These observations were acquired using the on-board Visual and Infrared Mapping Spectrometer (VIMS) and the application of this technique has been detailed in~\citet{Stewart2013, Stewart2015c, Stewart2015a}.
These images were recovered from occultation events at 9 ring edges as listed in Table~\ref{tab:KT}.
The opening angle between ring-plane and the line of sight to the star is quite shallow at -11$^\circ$, substantially increasing the opacity of the rings, and enhancing the edge sharpness.
These events came from three observations of the star passing behind the ring system which occurred within a one month period.
The relatively slow evolution which this star has previously exhibited~\citep{Tuthill2000} permits data from these temporally close observations to be used together in a single tomographic reconstruction.
The sampling resolution of these observations ranged from 25 to 63 mas with a mean of 39 mas, producing an image with a formal angular resolution slightly inferior to the aperture masking observations.
Due to \textit{Cassini}'s orbital geometry, the Position Angles ($P.A.$) of the occultations are clustered in two directions, which are separated by \tild35$^\circ$.
This angular diversity is relatively poor for image reconstruction, and results in a stretch to the image in the direction of the projections (orthogonal to the recovered spatial information), in this case approximately aligned in the north-south direction.
Kronocyclic tomographic imagery has been recovered in 4 broad spectral bands and accumulated into a single epoch.

\begin{table}
  \caption{\label{tab:KT}A list of sharp edges within Saturn's rings used in the image reconstruction using kronocyclic tomography.
  The names of edges are as defined by~\citet{Colwell2009} with identifiers from~\citet{French1993} where ‘IEG’ and ‘OER’ indicated the inner and outer edges, of gaps or rings respectively.
  The italicised \textit{i} or \textit{e} indicate if the event occurred during the ingress or egress of the entire occultation.
  The Position Angle (P.A.) shows the direction of occultation where the terrestrial celestial North is zero degrees and the angle increases toward the east.
  }
  \centering
  \begin{tabular}{c c r r}
\hline \hline
Date		&Edge			&$P.A.$	&$\theta_S (mas)$\\
\hline

03 Jun 2008	&A OER \textit{i}	&351.7	&62.65\\
		&B OER \textit{e}	&27.25	&61.28\\

10 Jun 2008	&A OER \textit{i}	&351.0	&49.95\\
		&Encke OEG \textit{i}&350.1	&49.99\\

02 Jul 2008	&Keeler OEG \textit{i}&349.0	&25.10\\
		&Keeler IEG \textit{i}&349.0	&25.00\\
		&Encke IEG \textit{i}&347.7	&25.02\\
		&Encke OEG \textit{e}&22.65	&26.05\\
		&Keeler OEG \textit{e}&21.67	&26.34\\
\end{tabular}

\end{table}

\section{Results}

This section presents reconstructed images, photometry, and additional supporting data which are then used to present the major observational findings of the paper.

\begin{figure}
 \centering
 \includegraphics[width=.99\columnwidth]{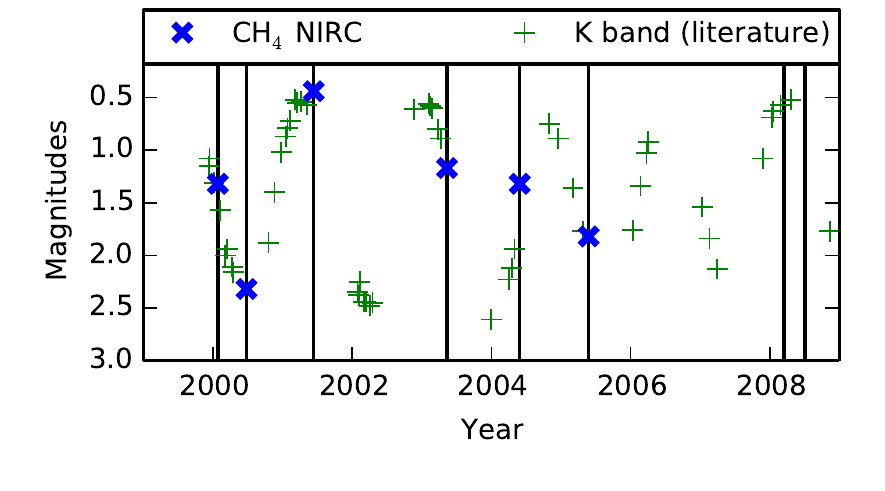}
 \caption{IRC+10216 Photometry: The green pluses show photometry recorded in K band by~\citet{Shenavrin2011}, whilst the blue crosses show photometric measurements for each of our NIRC epochs. 
 Our photometric measurements use NIRC's CH$_{4}$ filter which is approximately centred on K band.
 They were recorded concurrently with the aperture masking observations presented in Figure~\ref{fig:ch4}, and are consistent with the literature K band measurements.
 Each epoch is indicated by a black vertical line.}
 \label{fig:photometry}
\end{figure}

As a by product of the NIRC aperture masking observations, coincidental photometry was able to be recovered and is presented in Figure~\ref{fig:photometry}.
The blue crosses show the magnitude of the star in NIRC's CH$_4$ filter during the aperture masking observations.
The pass band of NIRC's CH$_4$ filter lies entirely within the broader K band, which is represented by green plus signs and comes from~\citet{Shenavrin2011}. 
The NIRC photometry presented here is shown to be consistent with the literature values, closely matching the observed fluctuations in brightness.

\subsection{NIRC maps: 2000-2005 in two colours}

\begin{figure*} 
\centering
  \def\svgwidth{1.7\columnwidth}
 \executeiffilenewer{ch4_maps_annotated.svg}{ch4_maps_annotated.pdf}%
 {inkscape -z -D --file=ch4_maps_annotated.svg %
 --export-pdf=ch4_maps_annotated.pdf --export-latex}%
 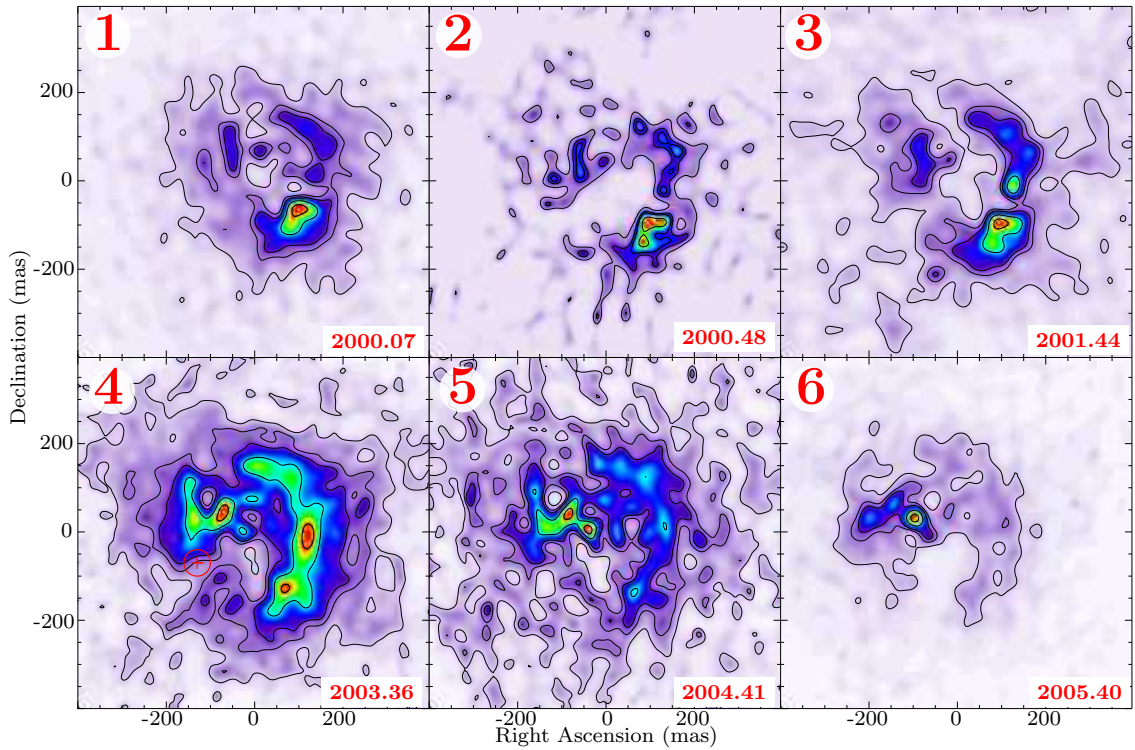%

  \caption{Six epochs between 2000 and 2005 observed with NIRC using the CH$_4$ filter.
  Contours are 2, 6, 10, 30, and 70\% of peak flux.
  The epochs are identified by the number in the top left corner of each map and correspond to the first six epochs in Table~\ref{tab:obs}.
  The decimal year of each epoch is given in the lower right of each panel.
  The red circle in panel 4 indicates the possible location of the stellar source based on polarimetry from~\citet{Murakawa2005} and is presented in closest epoch.
  All axes are equal and expressed in milliarcseconds with the vertical axis being declination and the horizontal axis being right ascension.
  North is up and East is the the left.
  }
  \label{fig:ch4} 
\end{figure*}

\begin{figure*}
 \centering
 \def\svgwidth{1.7\columnwidth}
 \executeiffilenewer{pahcs_maps_annotated.svg}{pahcs_maps_annotated.pdf}%
 {inkscape -z -D --file=pahcs_maps_annotated.svg %
 --export-pdf=pahcs_maps_annotated.pdf --export-latex}%
 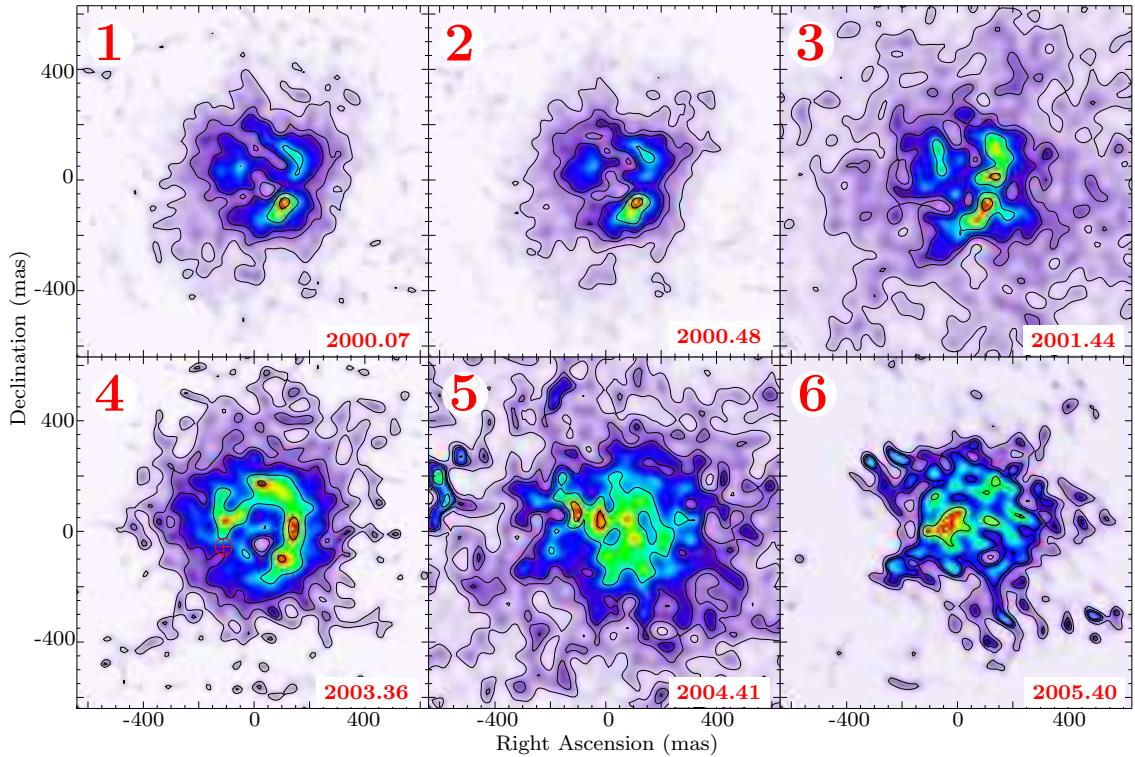%

 \caption{As with Figure~\ref{fig:ch4}, with NIRC's PAHCS filter.}
 \label{fig:pahcs}
\end{figure*}

IRC+10216 was observed by NIRC six times between 2000 and 2005 (listed in Table~\ref{tab:obs}) and photometry (shown in Figure~\ref{fig:photometry}) reveals that these epochs spanned almost three stellar brightness cycles.
The resulting maps observed at these epochs are presented in Figure~\ref{fig:ch4} for the CH$_4$ filter and in Figure~\ref{fig:pahcs} for the PAHCS filter.
Because the aperture masking technique is unable to provide astrometric positioning, maps from epochs two to six have been cross-correlated with their preceding epoch in order to ensure that features persisting longer than a single epoch align, and the slow evolution of the system can be monitored.

The first three epochs span a single cycle of the star's brightness oscillations and reveal a circumstellar environment generally similar to that discussed previously in the literature~\citep{Tuthill2000, Weigelt2002}.
The previously identified features are observed to continue moving apart.
The increasing separation between the `core' and the `Eastern Complex' is consistent with the rate of 17.8$\pm$1.9 mas/yr given by~\citet{Tuthill2000} with the separation at epoch 3 increasing to 230$\pm$8\,mas.
This is slightly above the predicted 228$\pm$4\,mas, but well within the uncertainties.
In epoch 3, the southern end of the North arm closest to the core of \textit{A} starts to brighten slightly in the CH$_4$ band and much more significantly in the redder PAHCS band.

Epoch 4 occurs approximately one stellar oscillation after epoch 3 and shows a remarkable dimming of \textit{A}.
An merger and increase in brightness is observed in the previously distinct features labelled as the Southern Component, the North Arm and the North-East Arm.
It is particularly significant that component \textit{A} ceases to be the dominant feature, which has not been observed at any earlier epochs for which high-resolution imagery has been recovered.
There is also a significant brightening if several knots in the Eastern Complex.

The \textit{A},\textit{B},\textit{C},\textit{D} nomenclature, already severely challenged by profound morphology changes up to 2005, now appears unusable and impossible to map onto present structures.
Over the last two epochs of this series the broad western structures fade quite rapidly as the knots in the Eastern Complex increase in brightness, eventually becoming the dominant feature.
The brightest of these appears to change places between epoch 4 and 6, with both visible in epoch 5.
Epoch 6 shows a clear dimming of the western arc until it is barely above the background noise level in the CH$_4$ map, and has lost most of its distinctive shape in the PAHCS band.

It is significant that there is no activity in the region around \textit{B} over all six of these epochs with the exception of some slight dimming into the level of the noise over the first two epochs.
The region around \textit{A} also approaches the background level over the last two epochs as it appears to fade from view, as does much of the North-East structure associated with \textit{C}.
The structures which had previously been detected around \textit{D} fade through the first three epochs and then undergo extensive brightening though the last three epochs of this series.
The apparent coming and going of these various features suggest that it is unlikely any of them are in any way permanent structures, rather they are most likely relatively short-lived transient features.

A relatively high level of background noise can be seen in the maps for some epochs, particularly epoch 5.
The poor quality of these maps is likely a consequence of the absence of any strong central core or brightest knot: noisy reconstructions are most notable et epochs with separated regions of similar intensity.
This is likely the result of the large, resolved structure giving very low visibility values everywhere.
The outcome is that the exact shape of specific features in this epoch are less reliable, although the general envelope and location of peaks appears robust.

\subsection{2008 CONICA and \Cassini observations}

\begin{figure*}
 \centering
 \def\svgwidth{1.99\columnwidth}
 \executeiffilenewer{conica_maps_annotated.svg}{conica_maps_annotated.pdf}%
 {inkscape -z -D --file=conica_maps_annotated.svg %
 --export-pdf=conica_maps_annotated.pdf --export-latex}%
 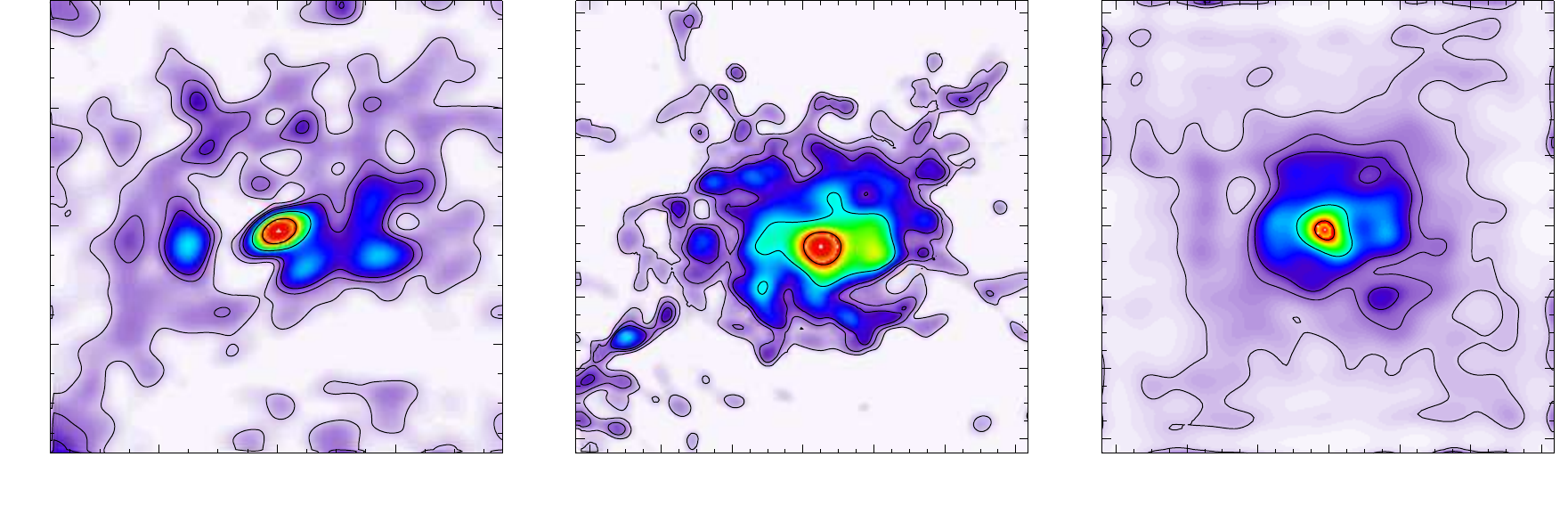%

 \caption{One epoch in March 2008 observed with CONICA using the IB\_2.24, NB\_3.74 and NB\_4.05 filters.
 Contours are 2, 6, 10, 30, and 70\% of peak flux.
 All axes are expressed in milliarcseconds with the vertical axis being declination and the horizontal axis being Right Ascension (R.A.).
 The scale of the first panel is different to that of the other two panels.
 The decimal year of of the observation is given in red in the lower right of the rightmost panel.
 North is up and East is the the left.}
 \label{fig:conica}
\end{figure*}

The CONICA epoch occurred more than three years after the last of the NIRC epochs, allowing the passage of more than 1.5 stellar brightness cycles in the interim.
IRC+10216 was observed by CONICA in March 2008 in three wavebands and the recovered maps are presented in Figure~\ref{fig:conica}.

\begin{figure*}
 \centering
  \includegraphics[width=1.6\columnwidth,clip=true,trim=25pt 20pt 4pt 48pt]{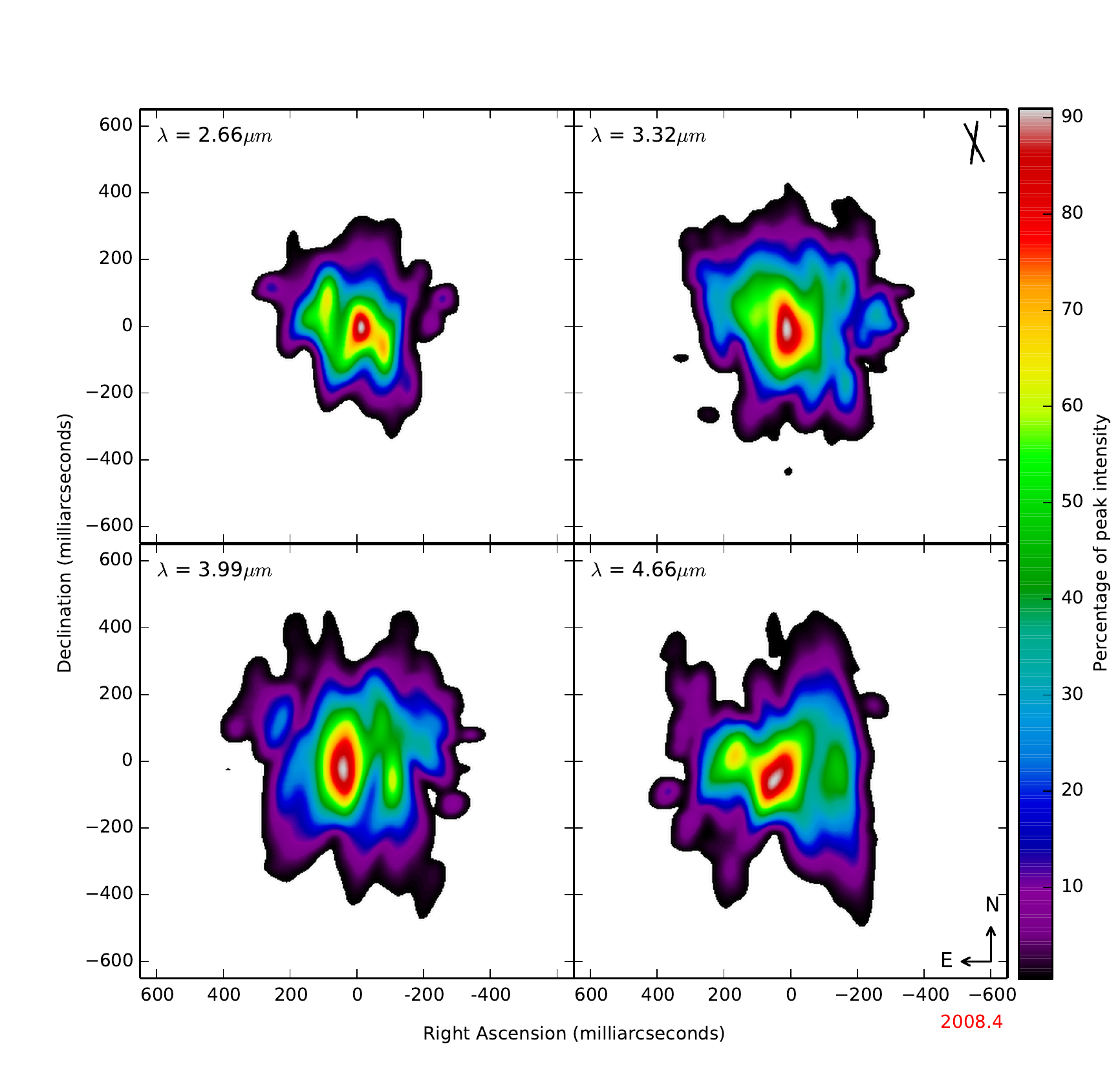}
 \caption{One epoch in June/July 2008 observed with \CassiniVIMS and recovered using kronocyclic tomography.
 Adjacent spectral channels were co-added into 4 broad spectral bands centred on $\lambda$~=~2.66, 3.32, 3.99, 4.66\,$\mu m$.
 The axes for all panels are identical, scaled as milliarcseconds in right ascension or declination.
 The centre wavelength is indicated in the top left of each panel.
 The angular diversity rose (defined in~\citet{Stewart2015a}) in the top right panel shows the occultation angles, and sampling resolution of the occultation events used in the recovery of these images (quantified in Table~\ref{tab:KT}).
 The decimal year of of the observation is given in red beneath the lower right panel.
 North is to the top and east is the the left.}
 \label{fig:cassini}
\end{figure*}

The maps reconstructed using kronocyclic tomography using \Cassini observations are shown in Figure~\ref{fig:cassini}.
These maps are generated from observations at 3 epochs within a single month, however, as shown in Table~\ref{tab:KT} the finest angular resolution spatial measurements all occurred in the final epoch.
Consequently, the finest spatial features in the recovered maps are most consistent with the final date of these epochs, whilst the earlier, lower resolution observations contribute more significantly to the  envelope.
The limited diversity of occultation angles produces an unrealistic stretch in aspect ratio over the features in the north-south direction which is evident when making a comparison to the temporally near CONICA epoch in Figure~\ref{fig:conica}.

The change in the overall appearance of the nebula from the later NIRC epochs to the CONICA and \Cassini epochs is striking, and it is not possible to confidently track the movement and evolution of existing features through to 2008.
These epochs reveal that the nebulosity around the star appears to continue to form bright knots of around the same spatial size (\tild50\,mas) as observed over the previous epochs and in the literature.
Unlike the NIRC pre-2006 epochs, the structure of the object appears to predominately lie along a line at \tild70$^\circ$ east of north.
All bands in both epochs show a larger central knot with a bright core which has strong knots on either side to the ENE and WSW.
A larger fainter envelope of radius approximately 300\,mas is observed to be asymmetrically filled, exhibiting an increased amount of flux in the northern half than the southern half in all wavelengths.
Unfortunately these two epochs are temporally too close to make any dependable measurements of the proper motions of the identified features.

\section{Discussion}

\subsection{Previous morphological models of IRC+10216}

Various qualitative and quantitative models for the nature of the inner parts of IRC+10216 have been proposed over the years based on observed structures.
In order to make sense of the reconstructed images published herein, it is necessary to examine the existing observations and models in the literature and consider how they apply to newer data.

Both~\citet{Weigelt1998} and~\citet{Haniff1998} published high resolution reconstructed images of the inner regions of the system.
They used a consistent naming scheme for the brightest four parts of the nebula, (\textit{A}, \textit{B}, \textit{C}, \textit{D} in order of decreasing brightness as discussed earlier).
This naming scheme has dominated the literature since, although there have been several additions or alterations to it as required by new observations and the identification of new, or reinterpreted features.
At this early stage both authors believed that the stellar photosphere was at least partially visible within the brightest component (\textit{A}) which they referred to as the 'central object'~\citep{Weigelt1998} or 'core'~\citep{Haniff1998}.
Weigelt suggested that the other components were knots within discrete dust layers, and Haniff supported the idea of a spherical dust envelope with wind-blown holes.

~\citet{Osterbart2000} revealed that the four brightest components were moving apart and that \textit{B}, \textit{C} and \textit{D} were dimming. 
They also published polarimetric imaging, however, this failed to pinpoint the relative position of the star itself.
They ultimately made the claim that the star was probably within or near component \textit{B}, although could not exclude the possibility the star was in the dark region between \textit{A} and \textit{B}, obscured by thick dust.

The location of the star was put back into component \textit{A} by~\citet{Tuthill2000} with higher resolution NIRC aperture masking observations.
Other structures were identified within the nebula, many of which are identified in Figure~\ref{fig:schematic}.
Again the identified features were found to be diverging.
Specifically, the North-East Arm (\textit{B},\textit{C}) was noted to be moving outward and increasing in length and the Eastern Complex (\textit{D}) was found to be moving eastward.
The bright core in the Southern Component (\textit{A}) was found to be close to the expected angular diameter of the star, supporting the claim that this was a direct view of the stellar photosphere.

In an ambitious endeavour to develop a two dimensional radiative transfer model describing the observed structure, \citet{Men'shchikov2001} countered that the star had to be within component \textit{B}.
The proposed model claimed to be able to reproduce observations at wavelengths across the near-infrared, but failed to identify the cause of most of the observed features, focusing instead solely on components \textit{A} and \textit{B}.
The model suggests that \textit{B} is direct, albeit extincted light from the star, and \textit{A} is one of a bipolar outflow cavity pair.
The model itself is highly complicated requiring tuning of over 20 parameters, many of which had little basis in previous measurements and were given large potential ranges.
There are also many assumptions of symmetry and shape with insufficient evidence to support them.
The model was extended to represent a larger sample of epochs and declared the previously published increasing separations of the components to be a ``pure projection effect"~\citep{Men'shchikov2002}.

\citet{Weigelt2002} found that the shape of the core of \textit{A} was continuously changing and becoming more elongated and that \textit{B} had faded and almost vanished completely by 2001.
In spite of this they still declare it to be the location of the star with the dimming explained away by a dramatic increase in mass-loss.

A lunar occultation observation found that both cores of \textit{A} and \textit{B} had approximately the same FWHM, allowing either component to potentially be the star.~\citep{Richichi2003}.
They also published an earlier occultation light-curve with only a single peak which had a FWHM consistent with component \textit{A} as observed in subsequent publications at other wavelengths.
This supported the notion that \textit{A} contained the stellar photosphere.
Multi-wavelength aperture masking observations also supported the idea that \textit{A} contained the photosphere~\citep{Tuthill2005}.
They made this argument based on several factors including \textit{A}'s angular size, K-band flux density, colour temperature, persistence over multiple epochs, observed dilution of photospheric absorption lines, level of mid-IR emission and magnitude of the proper motion between \textit{A} and \textit{B}.

Using polarimetric observations, \citet{Murakawa2005} claim to have located the star at 250$\pm$30\,mas east and 65$\pm$30\,mas south of the H band intensity peak on the 3rd of January 2003.
The peak intensity of the nebula has been previously shown to be consistent between H band and K band~\citep{Tuthill2005}, so we can use their offset from this peak to show their stellar position on our CH$_4$ map from our nearest observation.
This is identified in the nearest epoch, shown in panel 4 on Figures~\ref{fig:ch4} and~\ref{fig:pahcs}, with a red `+' surrounded by a red circle representing the polarimetric uncertainty.
Their stellar location is at the centre of the polarisation vector field, which they used this to exclude both \textit{A} and \textit{B} as likely stellar positions.
This position puts the star very close to the south of the existing structure labelled alternatively as \textit{D} or the Eastern Complex, however, in such a complex and asymmetric dusty environment there is no way to be certain that the polarisation field is well behaved and concentric on the star.

\citet{Weigelt2007} presented 12 epochs of observations spanning 1995 to 2005, overlapping with the presented NIRC data in Figures~\ref{fig:ch4} and~\ref{fig:pahcs}.
We find their maps to be broadly consistent with the observations presented here, showing the same divergence from the previously observed persistent structure over the period from 2003 to 2005.

\citet{Fonfria2014} observed IRC+10216 in several wave bands with CARMA (Combined Array for Research in Millimetre-wave Astronomy) at two epochs in 2011 and 2012.
They identified a range of molecular structures which are neither well aligned with each other nor with the existing structures in our 2008 epochs or the literature.

More recently,~\citet{Kim2015} published three epochs obtained with the Hubble Space Telescope.
These are of a lower resolution than the maps in this paper, similar to the maps produced by~\citet{Weigelt1998,Weigelt2007} and~\citep{Haniff1998}.
The first two of these, from 1998 and 2001, show the relatively stable but expanding structure identified previously and shown in the first 3 NIRC epochs in this paper.
Their third epoch, from 4th of June 2011, shows a vastly different map, with three close bright knots aligned along approximately 15$^\circ$.
The structure in this epoch shows no correlation to any previous observations.
They further identified a companion 500\,mas to the east of their brightest peak.
We find no evidence of this companion in any of our 8 near infrared epochs, nor is it to be found in any of the maps previously presented in the literature.
Rather than being a stellar companion appearing in only a single epoch in 23 years of observations (1989 in~\citet{Haniff1998} to 2012 in~\citet{Fonfria2014}), it seems much more likely that this feature is merely just another dusty knot in this highly dynamic environment.

\subsection{Prior models confront new imagery}

The previous models have been demonstrated to be both qualitatively and quantitatively inconsistent with newer observations.
These existing models have tended to focus on attempting to identify the location of the star itself and then claiming some of the observed features were part of some coherent circumstellar structure (e.g. bipolar) supporting the previous choice of stellar position.
With both of the most commonly argued-for stellar positions having faded from sight and being ruled out through polarimetry, these models have lost credibility.
Perhaps the most damaging aspect of the new observations from the perspective of the old models is the dramatic appearance of a new compact ``core'' several hundred milliarcseconds from both \textit{A} and \textit{B} in 2005.
It is worth noting that the possibility of the actual location of the star being concealed by thick dust was actually suggested by both~\citet{Osterbart2000} and~\citet{Tuthill2000}, but rejected in favour of \textit{B} and \textit{A} respectively.

With the benefit of hindsight and access to over two decades of observations in the literature, it is possible to rule out any of the existing features previously identified as the stellar photosphere.
We have shown that it is not possible to claim that the locations of any of the features are intrinsic to the underlying geometry of the system, or that they represent something fundamental about the location of the star or orientation of the inner nebula.
Instead we suggest that the size and asymmetric distribution of these knots can tell us about the time-evolving behaviour of the inner dust shroud.

The star driving the winds and forming these structures is expected to exhibit a uniform disc angular diameter of around 29\,mas~\citep{Menten2012}, consistent with period-mean density relationships scaling from other long period variables.
The various bright spots fall far outside the underlying photospheric radius, so must not simply be regions of unusually low opacity opening a line-of-sight to the photosphere itself.
The bright knots in the recovered images must be either regions of hot dust with high opacity, or windows in the dusty circumstellar material viewing hot material within, or some combination of both.
Either phenomenon has to be produced by some underlying process near the stellar photosphere and then propagate outward with the stellar wind.

The high brightness temperature (several thousand K) and persistence of the bright spots preclude explanations where hot material is simply carried out in the wind, because cooling times are measured in hours to days for clumps of moderate infrared optical depth.
Shock heating also appears to be an unlikely explanation, because it would require shock velocities comparable to the mean outflow velocity at radii where the wind has already likely reached near terminal velocity (\tild10\,stellar radii).
The requirement for the observations of a bright clump is a combination of a direct or near-direct path for radiation from the photosphere to reach the clump, and a low-opacity line of sight to the observer.

If the features are instead regions of lower opacity along the observer's line-of-sight, radiation from the hotter inner regions is able to be observed in a manner similar to the wind-blown holes proposed by~\citet{Haniff1998}.
Such holes would be formed by the turbulent stellar wind, exposing the inner regions of the nebula to an appropriately aligned observer.
The stellar photosphere is substantially smaller than the separation between the bright knots in the images ruling out the possibility of openings with a direct line-of-sight to the star itself.

\citet{Woitke2006a} produced 2D models of dust driven stellar winds from carbon-rich AGB stars.
They found that a range of instabilities, including Rayleigh-Taylor and Kelvin-Helmholtz, produced transient dusty arcs and knots in spite of the model's initial spherical symmetry.
This model predicted that such structures would move outwards over time, and that the dusty clouds are able to entirely obscure the line of sight to the star.
\citet{Woitke2008} demonstrated how these models predict that a fortuitously aligned observer can observe hotter material nearer to the star, albeit not the stellar photosphere, through such windows.
\citet{Freytag2008} presented complimentary 3D models of carbon-rich AGB stars which produce similar structures shown to be the result of inhomogeneous dust formation due to large convection cells originating beneath the photosphere.

Our preferred explanation is that the observed bright spots include both radiation-heated clumps in an inhomogeneous wind where both scattering and near-LTE emission processes operate, and opacity windows in a turbulent wind exposing hotter material nearer to the star.
Given the complexity of this system, a deeper understanding of this object requires 3D radiative transfer modelling of plausible inhomogeneous, clumpy winds.

\section{Conclusions}

We report high-resolution near-infrared imaging data which reveals morphological changes within the inner arcsecond of the nebula surrounding the evolved carbon star IRC+10216.
The dramatic nature of these changes supports the 2D models of~\citet{Woitke2006a} and the 3D models of~\citet{Freytag2008}.
These models generate a constantly evolving dusty environment which entirely envelops the star in an ever changing shroud, providing occasional brightened glimpses of the hot inner parts of the system.
We have demonstrated that none of the previously identified structures in IRC+10216's circumstellar environment are persistent.
These cannot therefore be representative of the alignment or position of the star within the dusty nebula, nor is it likely that models built on fundamental structural elements identified, such as bipolar axes, offer a useful way forward. 
Both features previously claimed to contain the stellar photosphere have faded from sight, and the polarimetric position determined by~\citet{Murakawa2005} is dubious as the complex and asymmetric circumstellar environment of IRC+10216 can't be assumed to produce a well-behaved, concentric polarisation field.
Instead, the star itself should be considered to be buried somewhere within its own constantly evolving clouds.

The high degree of polarisation observed by~\citet{Murakawa2005} confirms the importance of scattering in this nebula.
Polarimetric sparse aperture masking observations (such as is possible with NACO) would be able to differentiate between hot clumps (with a higher degree of polarisation) and opacity holes (with a lower degree of polarisation).

In order to make reliable claims about the stellar position within the nebula, images spanning a broad wavelength range at proximate epochs are necessary.
This will allow the registration of images from wavelengths where the starlight is certainly able to penetrate the nebula, such as \textit{N} band, with shorter bands where the photospheric emission is extincted.
Such observations will be possible with the upcoming MATISSE beam-combiner for the VLTI~\citep{Lopez2014,Kohler2014}, and will substantially advance our understanding of the complex inner regions of this intriguing object.

%

\bibliographystyle{mn2e} 
\bibliography{library}

\begin{thebibliography}{38}
\expandafter\ifx\csname natexlab\endcsname\relax\def\natexlab#1{#1}\fi

\bibitem[{Buscher(1994)}]{Buscher1994}
Buscher D.~F., 1994, in \prociau, p.~91

\bibitem[{Colwell {et~al}\mbox{.}(2009)Colwell, Nicholson, Tiscareno, Murray,
  French, \& Marouf}]{Colwell2009}
Colwell J.~E., Nicholson P.~D., Tiscareno M.~S., Murray C.~D., French R.~G.,
  Marouf E.~A., 2009, in Saturn from Cassini Huygens, Springer, pp. 375--412

\bibitem[{Decin {et~al}\mbox{.}(2011)Decin, Royer, Cox, Vandenbussche,
  Ottensamer, Blommaert, Groenewegen, Barlow, Lim, Kerschbaum, Posch, \&
  Waelkens}]{Decin2011}
Decin L. {et~al.}, 2011, \aap, 534, A1

\bibitem[{Fonfria {et~al}\mbox{.}(2014)Fonfria, Fernandez-Lopez, Agundez,
  Sanchez-Contreras, Curiel, \& Cernicharo}]{Fonfria2014}
Fonfria J.~P., Fernandez-Lopez M., Agundez M., Sanchez-Contreras C., Curiel S.,
  Cernicharo J., 2014, \mnras, 445, 3289

\bibitem[{French(1993)}]{French1993}
French R., 1993, \icarus, 103, 163

\bibitem[{Freytag \& H{\"{o}}fner(2008)}]{Freytag2008}
Freytag B., H{\"{o}}fner S., 2008, \aap, 483, 571

\bibitem[{Haniff \& Buscher(1998)}]{Haniff1998}
Haniff C.~A., Buscher D.~F., 1998, \aap, 334, L5

\bibitem[{Ireland(2006)}]{Ireland2006b}
Ireland M.~J., 2006, Proceedings of SPIE, 6268, 62681T

\bibitem[{Kastner \& Weintraub(1994)}]{Kastner1994}
Kastner J. H.~J., Weintraub D.~A., 1994, \apj, 434, 719

\bibitem[{Kim {et~al}\mbox{.}(2015)Kim, Lee, Mauron, \& Chu}]{Kim2015}
Kim H., Lee H.-G., Mauron N., Chu Y.-H., 2015, \apj, 804, L10

\bibitem[{K{\"{o}}hler {et~al}\mbox{.}(2014)K{\"{o}}hler, Ruge, Pott, Wolf,
  Jaffe, \& Henning}]{Kohler2014}
K{\"{o}}hler R., Ruge J.~P., Pott J.-U., Wolf S., Jaffe W., Henning T., 2014,
  in \procspie, Vol. 9146, p. 91461R

\bibitem[{Lacour {et~al}\mbox{.}(2011)Lacour, Tuthill, Ireland, Amico, \&
  Girard}]{Lacour2011}
Lacour S., Tuthill P.~G., Ireland M.~J., Amico P., Girard J., 2011, The
  Messenger, vol. 146, p. 18-23

\bibitem[{Le{\~{a}}o {et~al}\mbox{.}(2006)Le{\~{a}}o, Laverny, M{\'{e}}karnia,
  Medeiros, \& Vandame}]{Leao2006}
Le{\~{a}}o I.~C., Laverny P.~D., M{\'{e}}karnia D., Medeiros J. R.~D., Vandame
  B., 2006, \aap, 194, 187

\bibitem[{Lopez {et~al}\mbox{.}(2014)Lopez, Lagarde, Jaffe, Petrov,
  Sch{\"{o}}ller, Antonelli, Beckmann, Berio, Bettonvil, Glindemann, Gonzalez,
  Graser, Hofmann, Millour, Robbe-Dubois, Venema, Wolf, Henning, Lanz, Weigelt,
  Agocs, Bailet, Bresson, Bristow, Dugu{\'{e}}, Heininger, Kroes, Laun,
  Lehmitz, Neumann, Augereau, Avila, Behrend, van Belle, Berger, van Boekel,
  Bonhomme, Bourget, Brast, Clausse, Connot, Conzelmann, Cruzal{\`{e}}bes,
  Csepany, Danchi, Delbo, Delplancke, Dominik, van Duin, Elswijk, Fantei,
  Finger, Gabasch, Gay, Girard, Girault, Gitton, Glazenborg, Gont{\'{e}},
  Guitton, Guniat, {De Haan}, Haguenauer, Hanenburg, Hogerheijde, ter Horst,
  Hron, Hugues, Hummel, Idserda, Ives, Jakob, Jasko, Jolley, Kiraly,
  K{\"{o}}hler, Kragt, Kroener, Kuindersma, Labadie, Leinert, {Le Poole},
  Lizon, Lucuix, Marcotto, Martinache, Martinot-Lagarde, Mathar, Matter,
  Mauclert, Mehrgan, Meilland, Meisenheimer, Meisner, Mellein, Menardi, Menut,
  Merand, Morel, Mosoni, Navarro, Nussbaum, Ottogalli, Palsa, Panduro, Pantin,
  Parra, Percheron, Duc, Pott, Pozna, Przygodda, Rabbia, Richichi, Rigal,
  Roelfsema, Rupprecht, Schertl, Schmidt, Schuhler, Schuil, Spang, Stegmeier,
  Thiam, Tromp, Vakili, Vannier, Wagner, \& Woillez}]{Lopez2014}
Lopez Â. {et~al.}, 2014, The Messenger, 157, 5

\bibitem[{Matthews, Gerard \& {Le Bertre}(2015)Matthews, Gerard, \& {Le
  Bertre}}]{Matthews2015}
Matthews L.~D., Gerard E., {Le Bertre} T., 2015, \mnras, 449, 220

\bibitem[{Mauron \& Huggins(1999)}]{Mauron1999}
Mauron N., Huggins P.~J., 1999, \aap, 349, 203

\bibitem[{Mauron \& Huggins(2000)}]{Mauron2000}
Mauron N., Huggins P.~J., 2000, \aap, 359, 707

\bibitem[{Men'shchikov {et~al}\mbox{.}(2001)Men'shchikov, Balega,
  Bl{\"{o}}cker, Osterbart, \& Weigelt}]{Men'shchikov2001}
Men'shchikov A.~B., Balega Y., Bl{\"{o}}cker T., Osterbart R., Weigelt G.,
  2001, \aap, 368, 497

\bibitem[{Men'shchikov, Hofmann \& Weigelt(2002)Men'shchikov, Hofmann, \&
  Weigelt}]{Men'shchikov2002}
Men'shchikov A.~B., Hofmann K.~H., Weigelt G., 2002, \aap, 392, 921

\bibitem[{Menten {et~al}\mbox{.}(2012)Menten, Reid, Kamiński, \&
  Claussen}]{Menten2012}
Menten K.~M., Reid M.~J., Kamiński T., Claussen M.~J., 2012, \aap, 543, A73

\bibitem[{Murakawa {et~al}\mbox{.}(2005)Murakawa, Suto, Oya, Yates, Ueta, \&
  Meixner}]{Murakawa2005}
Murakawa K., Suto H., Oya S., Yates J.~A., Ueta T., Meixner M., 2005, \aap,
  436, 601

\bibitem[{Osterbart {et~al}\mbox{.}(2000)Osterbart, Balega, Bloecker,
  Men'shchikov, \& Weigelt}]{Osterbart2000}
Osterbart R., Balega Y., Bloecker T., Men'shchikov A.~B., Weigelt G., 2000,
  \aap, 357, 169

\bibitem[{Richichi, Chandrasekhar \& Leinert(2003)Richichi, Chandrasekhar, \&
  Leinert}]{Richichi2003}
Richichi A., Chandrasekhar T., Leinert C., 2003, \na, 8, 507

\bibitem[{Shenavrin, Taranova \& Nadzhip(2011)Shenavrin, Taranova, \&
  Nadzhip}]{Shenavrin2011}
Shenavrin V.~I., Taranova O.~G., Nadzhip A.~E., 2011, Astronomy Reports, 55, 31

\bibitem[{Sivia(1987)}]{Sivia1987}
Sivia D., 1987, PhD thesis

\bibitem[{Stewart {et~al}\mbox{.}(2013)Stewart, Tuthill, Hedman, Nicholson, \&
  Lloyd}]{Stewart2013}
Stewart P.~N., Tuthill P.~G., Hedman M.~M., Nicholson P.~D., Lloyd J.~P., 2013,
  \mnras, 433, 2286

\bibitem[{Stewart {et~al}\mbox{.}(2015{\natexlab{a}})Stewart, Tuthill,
  Nicholson, \& Hedman}]{Stewart2015c}
Stewart P.~N., Tuthill P.~G., Nicholson P.~D., Hedman M.~M.,
  2015{\natexlab{a}}, \mnras, Submitted

\bibitem[{Stewart {et~al}\mbox{.}(2015{\natexlab{b}})Stewart, Tuthill,
  Nicholson, Hedman, \& Lloyd}]{Stewart2015a}
Stewart P.~N., Tuthill P.~G., Nicholson P.~D., Hedman M.~M., Lloyd J.~P.,
  2015{\natexlab{b}}, \mnras, 449, 1760

\bibitem[{Thi{\'{e}}baut(2008)}]{Thiebaut2008}
Thi{\'{e}}baut E., 2008, in \procspie, pp. 70131I--70131I--12

\bibitem[{Tuthill {et~al}\mbox{.}(2010)Tuthill, Lacour, Amico, Ireland, Norris,
  Evans, Kraus, Lidman, Pompei, Paris, Cnrs, Xxx, Janssen, Observatory,
  Stewart, \& Kornweibel}]{Tuthill2010a}
Tuthill P.~G. {et~al.}, 2010, in \procspie, Vol. 7735, pp. 77351O--77351O--11

\bibitem[{Tuthill, Monnier \& Danchi(2005)Tuthill, Monnier, \&
  Danchi}]{Tuthill2005}
Tuthill P.~G., Monnier J.~D., Danchi W.~C., 2005, \apj, 624, 352

\bibitem[{Tuthill {et~al}\mbox{.}(2000{\natexlab{a}})Tuthill, Monnier, Danchi,
  \& Lopez}]{Tuthill2000}
Tuthill P.~G., Monnier J.~D., Danchi W.~C., Lopez B., 2000{\natexlab{a}}, \apj,
  543

\bibitem[{Tuthill {et~al}\mbox{.}(2000{\natexlab{b}})Tuthill, Monnier, Danchi,
  Wishnow, \& Haniff}]{Tuthill2000a}
Tuthill P.~G., Monnier J.~D., Danchi W.~C., Wishnow E.~H., Haniff C.~A.,
  2000{\natexlab{b}}, \pasp, 112, 555

\bibitem[{Weigelt {et~al}\mbox{.}(1998)Weigelt, Balega, Bl{\"{o}}cker,
  Fleischer, Osterbart, \& Winters}]{Weigelt1998}
Weigelt G., Balega Y., Bl{\"{o}}cker T., Fleischer A.~J., Osterbart R., Winters
  J.~M., 1998, \aap, 54, 51

\bibitem[{Weigelt {et~al}\mbox{.}(2002)Weigelt, Balega, Bl{\"{o}}cker, Hofmann,
  Men'shchikov, \& Winters}]{Weigelt2002}
Weigelt G., Balega Y., Bl{\"{o}}cker T., Hofmann K.~H., Men'shchikov A.~B.,
  Winters J.~M., 2002, \aap, 392, 131

\bibitem[{Weigelt {et~al}\mbox{.}(2007)Weigelt, Balega, Hofmann, Men'shchikov,
  Murakawa, \& Schertl}]{Weigelt2007}
Weigelt G., Balega Y., Hofmann K.~H., Men'shchikov A.~B., Murakawa K., Schertl
  D., 2007, in Why Galaxies Care About AGB Stars: Their Importance as Actors
  and Probes, Vol. 378, p. 349

\bibitem[{Woitke(2006)}]{Woitke2006a}
Woitke P., 2006, \aap, 452, 537

\bibitem[{Woitke(2008)}]{Woitke2008}
Woitke P., 2008, \prociau, 4, 229

\end{thebibliography}

\bsp

\label{lastpage}

\end{document}